\documentclass[conference]{IEEEtran}
\usepackage{amssymb,latexsym,euscript,array,rotating,graphics}
\usepackage{url}
\usepackage{soul} 

\usepackage{picinpar,graphicx}

\usepackage{alltt}
\usepackage{pgf}
\usepackage{epsfig}

\usepackage{array}

\usepackage{amsthm}

\usepackage{color}
\definecolor{color-alois}{rgb}{0,0,1}
\newcommand{\alois}{\color{color-alois}}
\newcommand{\aloisbox}[1]{\textcolor{color-alois}{#1}}
\newcommand{\black}{\color{black}}
\newcommand{\OK}[1]{\textcolor{red}{#1}}
\newcommand{\PCH}[1]{\textcolor{green}{#1}}
\definecolor{color-fred}{rgb}{1,0.2,0}
\newcommand{\fred}[1]{\textcolor{color-fred}{#1}}
\newcommand{\replace}[2]{\st{#1} #2}

\renewcommand{\alois}{}
\renewcommand{\aloisbox}{}
\renewcommand{\black}{}
\renewcommand{\OK}{}
\renewcommand{\PCH}{}
\renewcommand{\fred}{}
\renewcommand{\replace}[2]{#2}

   \newtheoremstyle{example}
   	 {}
   	 {}%
     {\it}
     {}
     {\bfseries}
     {}
     {5pt}
     {\thmname{#1} \thmnumber{#2} -- \textbf{\thmnote{#3}}.}

   \theoremstyle{example}
   \newtheorem{example}{Example}

\newtheorem{proposition}{Proposition}

\usepackage[cmex10]{amsmath}

\usepackage{mdwmath}
\usepackage{mdwtab}

\usepackage{tikz}
\usepackage{pdfpages}
\usetikzlibrary{arrows}
\usetikzlibrary{automata}
\usetikzlibrary{er}
\usetikzlibrary{matrix}

\usepackage{url}

\def \sp {\hspace*{0.6cm}}

\begin{document}
%
\title{Random Grammar-based Testing for \\
Covering All Non-Terminals} 


\author{\IEEEauthorblockN{Alo\"is Dreyfus, Pierre-Cyrille H\'eam and Olga Kouchnarenko}
\IEEEauthorblockA{FEMTO-ST - CNRS UMR6174 - Universit\'e de Franche-Comt\'e - INRIA CASSIS\\
16 route de Gray - 25030 Besan\c{c}on, France\\
Email: firstname.\aloisbox{last}name@femto-st.fr}

}


%


\maketitle

\begin{abstract}
In the context of software testing, generating complex data inputs is
frequently performed using a grammar-based specification. For combinatorial
reasons, an exhaustive generation of the data -- of a given size -- is
practically impossible, and most approaches are either based on random
techniques or on coverage criteria. In this paper\fred{,} we show how to combine these
two techniques by biasing the random generation in order to optimise the
probability of satisfying a coverage criterion.
\end{abstract}


\alois \section{Introduction} \black
\label{sec-intro} \black

\subsection{Motivation}

Producing trusted software is a central issue in software
engineering.  Testing remains an inescapable step to ensure software
quality. In reaction to the limitations of manual testing, recent years have seen 
a rise in the research interest for systematic testing frameworks grounded in theory.
Random testing is a natural approach, empirically known to detect many kinds of bugs. 
However, by definition, low-probability behaviours cannot be adequately tested in that way.
Conversely, non-random testing tends to focus on a few edge cases of particular interest to the tester, at the expense of all others.
Indeed, it can  cover various behaviour\fred{s}, but their choice depends on tester's priorities
and\fred{,} in general\fred{,} each behaviour is tested in a unique way.

In~\cite{DBLP:journals/sttt/DeniseGGLOP12}, it is explained how to
bias a uniform random testing approach \fred{\replace{through}{using}} constraints given by a coverage
criterion, in order to optimise the probability of \fred{\replace{covering}{satisfying}} this
criterion.  The technique is developed for path generation in a graph. 
\aloisbox{The contribution of the \fred{present} paper}  \OK{consists in enriching this approach with 
a coverage criterion on \aloisbox{non-terminal} 
symbols of the grammar, allowing the user} 
to apply it to grammar-based testing.

\alois \subsection{Related Work} \black
 
Grammar-based testing is frequently used for generating structured inputs,
\fred{\replace{like}{as}} in~\cite{Purdom} for parser testing or in~\cite{ASTGen} to test
refactoring engines (program transformation software). Systematic
combinatorial approaches~\cite{DBLP:conf/kbse/CoppitL05} lead to a huge
number of sequences, and symbolic approaches are frequently
preferred~\cite{DBLP:conf/pts/LammelS06,DBLP:conf/kbse/MajumdarX07,DBLP:conf/pldi/GodefroidKL08}.
In~\cite{SEFM}, a generic tool for generating \fred{\replace{tests}{test data}} from grammars has been 
proposed. This tool does not provide any random feature but  is based on rule
coverage algorithms and techniques, as defined
in~\cite{Purdom,DBLP:conf/fase/Lammel01,DBLP:conf/compsac/ZhengW09,DBLP:conf/sle/AlvesV08}.

\fred{\replace{Random generation based approaches for testing}{Random test generation techniques}}
 -- initially
proposed in~\cite{802530,Ham94} -- are frequently used for practical
reasons, as in~\cite{DART,jartege,DBLP:journals/sttt/DeniseGGLOP12}.
Combining random generation and grammar-based testing is explored 
in~\cite{mckenzie,DBLP:journals/siamcomp/HickeyC83,DBLP:journals/spe/Maurer92,Seed,DBLP:journals/entcs/DadeauLH09,DBLP:conf/tap/HeamM11}%
\aloisbox{, without} \OK{exploiting any coverage criteria}, \fred{or using an isotropic random walk as in~\cite{DBLP:conf/icst/EnderlinDGB12}}.

\alois \subsection{Layout} 

Section~\ref{sec-background} presents the notions and \fred{\replace{the }{}}notations used in this paper. \black
Section~\ref{sec-RTCC} explains how to optimise random testing
\fred{\replace{relatively to a}{to satisfy a given}} coverage criterion. The theoretical contributions are
provided in Section~\ref{sec-proba}, \fred{\replace{where it is shown}{which shows}} how to use \fred{\replace{the}{this}}
technique \alois to optimise the coverage of non-terminal symbols \black in a grammar-based testing context.
An illustrating example is developed in Section~\ref{sec-running}.
\fred{Finally, Section~\ref{sec-conclusion} concludes.}

\alois \section{Formal Background} \black
\label{sec-background} \black

\subsection{Context-free Grammars and Random Generation}
In this paper, the \fred{\replace{cardinal}{cardinality}} of a finite set $\mathcal{S}$ is denoted $|\mathcal{S}|$.

\paragraph{Context-free Grammars}
 A {\it context-free grammar} is a tuple $G=(\Sigma,\Gamma,S_0,R)$, where
 $\Sigma$ and $\Gamma$ are disjoint finite alphabets, $S_0\in \Gamma$ is the
 initial symbol, and $R$ is a finite subset of $\Gamma\times (\Sigma\cup
 \Gamma)^*$.
\alois The elements of  $\Sigma$ are called {\it terminal symbols}, and the elements of $\Gamma$ are called {\it non-terminal symbols}. \black
An element $(X,u)$ of $R$ is called {\it a rule} of the grammar
 and is frequently denoted $X\to u$. A word
 $w\in (\Sigma\cup \Gamma)^*$ is a {\it successor} of $v \in (\Sigma\cup
 \Gamma)^*$ for the grammar $G$ if there exist $v_0\in \Sigma^*$,
 $v_1,v_2\in (\Sigma\cup \Gamma)^*$, $S\in \Gamma$ such that $v=v_0Sv_1$ and
 $w=v_0v_2v_1$ and $S\to v_2\in R$. 
A {\it complete derivation}\footnote{As $v_0\in \Sigma^*$, this derivation is obviously a left-most derivation.} of the
 grammar $G$ is a finite sequence $x_0,\ldots,x_k$ of words of $(\Sigma\cup
 \Gamma)^*$ such that $x_0=S_0$, $x_k\in \Sigma^*$ and for every $i$,
 $x_{i+1}$ is a successor of $x_i$.
A {\it derivation tree} of $G$ is a finite tree whose internal nodes are
label\aloisbox{l}ed by letters of $\Gamma$, whose leaves are labelled by elements of
$\Sigma\cup\{\varepsilon\}$, whose root is labelled by $S_0$ and satisfying:
if a node is label\aloisbox{l}ed by $X\in \Gamma$ and  its children are labelled by 
$\alpha_1,\ldots,\alpha_k$ (in this order), then either 
$\alpha_1=\varepsilon$ and $k=1$, or all the $\alpha_i$'s are in $\Gamma\cup
\Sigma$ and $(X,\alpha_1\ldots\alpha_k)\in R$. 
The size of a derivation tree is \fred{given by} the number of \fred{\replace{its}{tree}} nodes.

\begin{example}[Context-free grammar]\label{exe:grammar}
Let us consider \fred{\replace{for instance }{}}the grammar $G=(\{a,b\},\{S,T\},S,R)$, with
$R=\{S\to Tb,S\to aSb,T\to \varepsilon\})$. The sequence $S, aSb, aTbb, abb$
is a complete derivation of the grammar. The associated derivation tree is 

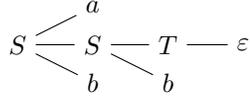
\begin{center}
\begin{tikzpicture}
\node (S) at (0,0) {$S$};
\node (S2) at (1,0) {$S$};
\node (a) at (1,0.5) {$a$};
\node (b) at (1,-0.5) {$b$};
\node (b2) at (2,-0.5) {$b$};
\node (T) at (2,0) {$T$};
\node (e) at (3,0) {$\varepsilon$};

\path[] (S) edge[above] node {} (S2);
\path[] (S) edge[above] node {} (a);
\path[] (S) edge[above] node {} (b);
\path[] (S2) edge[above] node {} (b2);
\path[] (S2) edge[above] node {} (T);
\path[] (T) edge[above] node {} (e);r
\end{tikzpicture}
\end{center}
\end{example}

Note that there is a bijection between the set of complete
derivations of a grammar and the set of derivation trees of this grammar.
For a context-free grammar $G$, $E_n(G)$ denotes the number of derivation
trees of $G$ with $n$ nodes. A derivation tree {\it \aloisbox{cover}s} an element $X$
of $\Gamma$ if at least one of its nodes is labelled by $X$.
For instance, for the tree in Example~\ref{exe:grammar}, 
 \aloisbox{the elements $S$ and $T$ are covered since they \OK{appear} in the~derivation~tree.}

\paragraph{Uniform Random Generation}

The present issue is, given a positive integer and a context-free
grammar, to compute randomly with a uniform distribution a derivation
tree of size $n$ of this grammar. We will briefly explain here how to
tackle this problem by using well-known counting
techniques~\cite{flajolet}. Notice that more advanced techniques allow
a faster computation, like in~\cite{DBLP:journals/tcs/DeniseZ99}.

\OK{As usual, }\aloisbox{the non-terminals symbols} \PCH{are denoted by} \OK{capital letters.}
Given a context-free grammar $G=(\Sigma,\Gamma,S_0,R)$,  \aloisbox{a \OK{non-terminal} symbol $X$ in $\Gamma$, and a positive integer $i$,}  \OK{the number of derivation trees of  size $i$  generated by} $(\Sigma,\Gamma,X,R)$  \OK{ is denoted  by $x(i)$,  i.e.,} \PCH{using} \OK{the corresponding lowercase~letter.}

Given a positive integer $n$,
for each symbol $S\in \Gamma$, the sequence of positive
integers $s(1),\ldots,s(k),\ldots$ is introduced.
The recursive computation of these $s(i)$'s is as follows.  For each
strictly positive integer $k$ and each rule $r=(S,w_1S_1\linebreak[4]\ldots
w_nS_nw_{n+1})\in R$, with $w_j\in \Sigma^*$ and $S_i\in \Gamma$, let
\fred{us} set
$$
\begin{cases}
\beta_r=1+\sum_{i=1}^{n+1} |w_i|\\
\alpha_r(k)=\sum_{i_1+i_2+\ldots+i_n=k-\beta_r}\prod_{j=1}^{j=n}s_j(i_j)
 \quad
\text{if } n\neq 0\\
\alpha_r(k)=0 \quad \text{if } n= 0 \text{ and } k\neq \beta_r\\
\alpha_r(\beta_r)=1 \quad \text{if } n= 0.
\end{cases}
$$
It is known~\cite[Theorem~I.1]{flajolet} that 
$s(k)=\sum_{r\in R\cap (S\times (\Sigma\cup\Gamma)^*)} \alpha_r(k).$

Since, by hypothesis, there is no rule of the form $(S,T)$ in $R$, with
$S,T\in \Gamma$, all $i_j$'s involved in the definition of $\beta_r$ are
strictly less than $k$. This way, the $s(i)$'s can be recursively computed.
Consider for instance the grammar
 $(\{a,b\},\{X\},X,\{r_1,r_2,r_3\})$ with
$r_1=(X,XX)$ $r_2=(X,a)$ and $r_3=(X,b)$.
One has $\beta_{r_1}=1+0=1$,
$\beta_{r_2}=1+1=2$, $\beta_{r_3}=1+1=2$. Therefore 
$x(k)=\sum_{i+j=k-1}x(i)x(j)$ if $k\neq 2$, and 
$x(2)=1+1+\sum_{i+j=2-1}x(i)x(j)=2$, otherwise. 
It follows that $x(1)=0$, $x(2)=2$, $x(3)=x(1)x(1)=0$,
$x(4)=x(1)x(2)+x(2)x(1)=0$, $x(5)=x(2)x(2)=4$, etc.
The two derivation trees of size 2 are $\substack{X\\ | \\a}$ and
$\substack{X\\ | \\b}$. The four derivation trees of size 5 are the trees of
the form $\substack{X\\ /\backslash \\{Z_1}\; Z_2}$ where both $Z_1$ and $Z_2$
are derivation trees of size 2.
 
In order to generate derivation trees of size $n$, all $s(i)'s$, for $S\in
\Gamma$ and $i\leq n$, have to be computed with the above method. This can
be performed in polynomial time. Afterwards, the random generation is done 
recursively using the \fred{\replace{Random Generation}{given}} algorithm in Fig.~\ref{fig:algo}.

\begin{figure}
\noindent
{\bf Random Generation}\\
{\bf Input:} $G=(\Sigma,\Gamma,X,R)$ a context-free grammar, $n$ a strictly
positive integer.\\
{\bf Output:} a derivation tree $t$ of $G$ size $n$.
\hrule
\vspace{0.1cm}
\noindent
{\bf Algorithm:} 

\noindent
\sp 1. Let $\{r_1,r_2,\ldots,r_\ell\}$ be set of the elements of $R$ whose first
  element is~$X$.\newline
\sp 2. {\bf If} $\sum_{j=1}^{j=\ell}\alpha_{r_j}(n)=0$, {\bf then} {\bf return} ``Exception''.\newline
\sp 3. {\bf Pick} $i\in\{1,\ldots,\ell\}$ with probability
$Prob(i=j)=\frac{\alpha_{r_i}(n)}{\sum_{j=1}^{j=\ell}\alpha_{r_j}(n)}.$\newline
\sp 4. Let $r_i=(X,Z_1\ldots Z_k)$, with $Z_j\in \Sigma\cup\Gamma$.\newline
\sp 5. Root symbol of $t$ is $X$.\newline
\sp 6. Children of $t$ are $Z_1,\ldots,Z_k$ in this order.\newline
\sp 7. Let $\{i_1,\ldots,i_m\}=\{j\mid Z_j\in \Gamma\}.$\newline
\sp 8. {\bf Pick} $(x_1,\ldots,x_m)\in \mathbb{N}^m$ such that
$x_1+\ldots+x_m=n-\beta_{r_i}$ with probability

$$Prob(x_1=\ell_1,\ldots,x_m=\ell_m)=\frac{\prod_{j=1}^{j=m}z_{i_j}(\ell_j)}{\alpha_{r_i}(n)}.$$  \newline
\sp 9. For each $i_j$, the $i_j$-th sub-tree of $T$ is obtained by running the
{\bf Ran\-dom Generation} algorithm on $(\Sigma,\Gamma,Z_{i_j},R)$ and
$\ell_j$.\newline
\sp 10. {\bf Return } $t$.\newline
\smallskip
\caption{Random Generation algorithm}\label{fig:algo}
\end{figure}

It is known~\cite{flajolet} that this algorithm provides a uniform
generation of derivation trees of size $n$, i.e. each derivation tree occurs
with the same probability. Note that an exception is \fred{\replace{returned}{raised}} at Step 2 if
there is no element of the given size.   For the example presented before,
there is no element of size~3, then it is impossible to generate a derivation tree of size~3.
Running the algorithm on this example with $n=2$,   one
consider\fred{s} at Step 1 the set  $\{r_1,r_2,r_3\}$ since all these rules have $X$ as left
element. Since $\alpha_{r_1}(2)=0$,   $\alpha_{r_2}(2)=1$,
$\alpha_{r_3}(2)=1$, at  Step 3 the probability that $i=1$ is \fred{\replace{null}{0}}, 
 the probability that $i=2$ is $1/2$ and  the probability that $i=3$ is
 $1/2$. If $i=2$ is picked, the generated tree has $X$ as root symbol
 and $a$ as unique child. Running the algorithm on this example with $n=3$
 stops at Step 2 since there is no tree of size $3$.  When running the
 algorithm on this example with $n=5$, the set $\{r_1,r_2,r_3\}$ is
 considered at Step 1. Since  $\alpha_{r_1}(5)=4$,   $\alpha_{r_2}(5)=0$,
$\alpha_{r_3}(5)=0$, $i=1$ is picked with probability 1. Therefore, the tree 
has $X$ as root symbol, and its two children are both labelled by $X$.
Therefore, at Step 7, the considered set is $\{1,2\}$. At Step 8, one has
$n-\beta_{r_1}=5-1=4$. The probability that $i_1=1$ and $i_2=3$ is 0
since $x(1)=0$.  Similarly, the probability that $i_1=3$ and $i_2=1$ is 0
too. Now the probability that  $i_1=2$ and $i_2=2$ is 1. Afterwards the algorithm
is recursively executed on each child with $n=2$: each of the 4 trees is
chosen with probability $1/4$.

\section{Mixing Random Testing and Coverage Criteria}
\label{sec-RTCC}

In a context of functional testing, the strength of random testing is
to \aloisbox{quickly provide} many different \fred{\replace{tests}{test data}}, 
\aloisbox{for each behaviour of the system.} 
Moreover, these \fred{\replace{tests}{test data}} are
independent of the choices of the test designer, and consequently they
can catch problem (s)he did not anticipate. For instance, fuzz testing
is particularly relevant for testing security
requirements~\cite{DBLP:conf/pldi/GodefroidKL08}. However, random
testing can miss an important behaviour occurring with a very small
probability. To exploit the advantages of both random testing and
deterministic testing, a solution \fred{\replace{would be}{is}} to combine random
generation and coverage criteria.

The general schema for this combination, as described
in~\cite{DBLP:journals/sttt/DeniseGGLOP12}, is the following:
considering a random generation algorithm of \fred{\replace{tests}{test data}} of size $n$ and a
coverage criteria $C$ (each element of $C$ is or is not covered by
each possible test), the goal is to use the generation algorithm $N$
times in order to optimise the probability of covering all elements of
$C$.  For each element $e\in C$, we denote by $p_{e,n}$ the
probability that a generated test of size $n$ covers $e$.  One can
easily check that 
\PCH{generating}  $N$ \fred{\replace{tests}{test data}}
independently of $C$ provides a probability of covering $C$ 
 of $1-(1-p_{\rm min})^N$, where $p_{\rm
  min}=\min_{e\in C}\{p_{e,n}\}$. 
\PCH{This probability is the way to measure the quality of the testing
  approach, relatively to~$C$.}
 A better way is to repeat $N$ times
the following procedure:
\begin{enumerate}
\item Pick at random an element $e\in C$ with a probability $\pi_e$, and
\item Generate uniformly a test of size $n$ covering $e$. 
\end{enumerate}
This procedure requires \fred{\replace{knowing}{to know}} how to uniformly generate a test of size
$n$ covering a given element, and to choose the probabilities \aloisbox{$\pi_e$}'s to
optimise the probability of covering all elements of $C$. 

Following~\cite{DBLP:journals/sttt/DeniseGGLOP12}, the optimisation requires
solving the following constraint system: maximise $p$ satisfying
$$
\begin{cases}
p\leq \sum_{e\in C} \pi_e \frac{p_{e,f,n}}{p_{e,n}}\text{ for all }
f\in C\\
\sum_{e\in C} \pi_e=1
\end{cases}
$$ where $p_{e,f,n}$ is the probability that a randomly generated test of size
$n$ covers both $e$ and $f$. This linear programming problem can be solved
in an efficient way, using simplex-like approaches. 

\fred{\replace{To sum up}{In summary}}, in order to combine random testing and a coverage criterion,
it is required to solve a constraint system and to know 1) how to randomly
generate a test of a given size covering a given element,  2) how to
compute the $p_{e,n}$'s; and 3) how to compute the $p_{e,f,n}$'s.

The rest of the paper is dedicated \aloisbox{to} the problem of the random
generation of execution trees of a grammar, with the coverage
criterion {\it All  \fred{\replace{rewriting}{non-terminal}} symbols}. More precisely, given a grammar
$G=(\Sigma,\Gamma,S_0,R)$, the coverage criterion being $\Gamma$, a
test of size $n$ being a derivation tree of $G$ of size $n$, we say
that $X\in \Gamma$ is covered by a test if the derivation tree \aloisbox{cover}s~$X$.

\section{Computing $p_{X,n}$ and  $p_{X,Y,n}$}\label{sec-proba}

In this section $G=(\Sigma,\Gamma,S_0,R)$ is a context-free grammar. 
We denote by $E_{n}(G)$ the set of derivation trees of size $n$ of $G$. 
We respectively denote by $E_{X,n}(G)$ \aloisbox{and} $E_{X,Y,n}(G)$
 the set of derivation trees of size $n$ of $G$ \aloisbox{cover}ing
 $X$\aloisbox{, and cover}ing both $X$ and $Y$
. 

Let $p_{X,n}$ be the probability of a randomly generated derivation
tree of size $n$ to \aloisbox{cover} $X$.  Clearly, if $E_{n}(G)$ is empty then
$p_{X,n}=0$ [resp. $p_{X,Y,n}=0$]. Otherwise,
$p_{X,n}=\frac{|E_{X,n}(G)|}{|E_n(G)|}$ [resp.
$p_{X,Y,n}=\frac{|E_{X,Y,n}(G)|}{|E_n(G)|}$].

Therefore, computing the probability
\fred{\replace{needed to solve the linear constraint program defined in Section~\ref{sec-RTCC},}
{defined in Section~\ref{sec-RTCC} -- needed to solve the linear constraint program  -- }}
reduces to the computation of the cardinality of sets $E_{X,n}(G)$ and
$E_{X,Y,n}(G)$.

\subsection{Computing  $|E_{X,n}(G)|$ and  $|E_{X,Y,n}(G)|$}

To compute $|E_{X,n}(G)|$, we build a grammar $G_X$ such that $E_n(G_X)$
and $E_{X,n}(G)$ are in bijection (and therefore have the same number of
elements).

For every $w\in (\Gamma\cup\Sigma)^*$, $[w]_0$ is recursively defined by:
$[\varepsilon]_0=\varepsilon$, $[Zw]_0=(Z,0)[w]_0$ (with $Z\in \Gamma$) and
$[aw]_0=a[w]_0$ (with $a\in \Sigma$). Intuitively, $[w]_0$ is obtained from $w$
by changing each letter of $w$ in $\Gamma$ by the corresponding pair with
$0$ as second element. For instance, with the grammar of
Example~\ref{exe:grammar}, one has $[aSbbT]_0=a(S,0)bb(T,0)$.
For every $w\in (\Gamma\cup\Sigma)^*$, $[w]_2$ is defined exactly in the
same way, changing all $0$'s by~$2$'s.

For every $w\in (\Gamma\cup\Sigma)^*$, $\{w\}_{1,2}$ is defined as the
set of words $w^\prime\in (\Sigma\cup\Gamma\times\{1,2\})^*$ obtained
from $w$ by \fred{\replace{unchanging the letters in $\Sigma$, }{}}
replacing occurrence
of each letter $Z$ of $\Gamma$ either by $(Z,1)$ or by $(Z,2)$, with
the restriction that at least one is replaced by $(Z,1)$.
\fred{The letters in $\Sigma$ remain unchanged.}
For instance\fred{,} if $w=aSbT$, then
$\{w\}_{1,2}=\{a(S,1)b(T,1),a(S,2)b(T,1),a(S,1)b(T,2)\}$. Notice that
if $w\in \Sigma^*$ then $\{w\}_{1,2}=\emptyset$ since the constraint
is not satisfied.

 Let $G_X=(\Sigma,\Gamma\times\{0,1,2\},(S_0,1),R_X)$
where $R_X=R_0\cup R_1\cup R_1^\prime\cup R_2$ with:
\begin{itemize}
\item $R_0=\{(Z,0)\to [w]_0\mid Z\to w\in R\}$,
\item $R_1=\{(Z,1)\to w^\prime\mid Z\neq X\text{ and }\exists Z\to w\in R\text{ such that }
  w^\prime\in \{w\}_{1,2}\}$,
\item $R_1^\prime=\{(X,1)\to [w]_0\mid X\to w\in R\}$,
\item $R_2=\{(Z,2)\to [w]_2\mid Z\to w\in R\text{ and }Z\neq X\}$.
\end{itemize}

Intuitively, adding the value $0$ to a symbol in $\Gamma$ means that
if this rule is \fred{\replace{fired}{used}}, there exists an \fred{occurrence of} $X$ at an upper position in the
derivation tree. Adding the value $1$ to a symbol in $\Gamma$ means
that there is no \fred{occurrence of} $X$ at an upper position, but there exists an \fred{occurrence of} $X$ at
this or a lower position in the derivation tree.  The value $2$ means
there is no \fred{occurrence of} $X$ appearing in the tree at an upper or lower position.

\begin{example}[$G_X$]\label{exe-GX}
  Consider the grammar $G=(\{a,b\},\{S,T,X\},S,R)$ with $R=\{S\to
  SS,S\to aT,S\to Xb,T\to aa, X\to TX, X\to b\}$. The grammar $G_X$
  has the set of rules as follows: $\{(S,0)\to (S,0)(S,0), (S,0)\to
  a(T,0), (S,0)\to (X,0)b, (T,0)\to aa, (X,0)\to b,(X,0)\to
  (T,0)(X,0)\} \cup\{ (S,1)\to (S,1)(S,1), (S,1)\to (S,1)(S,2),
  (S,1)\to (S,2)(S,1), (S,1)\to a(T,1), (S,1)\to (X,1)b\} \cup
  \{(X,1)\to b, (X,1)\to (T,0)(X,0)\} \cup \{(S,2)\to (S,2)(S,2),
  (S,2)\to a(T,2), (S,2)\to (X,2)b, (T,2)\to aa\}$.
\end{example}

\begin{proposition}[Bijection]\label{prop:phiX}
There exists a bijection between  $E_n(G_X)$ and $E_{X,n}(G)$.
\end{proposition}

\noindent Example~\ref{exe:proof} illustrates several elements of \fred{\replace{this}{the following}} proof. 
\begin{proof}
 Let $\varphi$ be the function from $(\Gamma\times\{0,1,2\}\cup \Sigma)^*$
 into $(\Gamma\cup \Sigma)$ inductively defined by:
 $\varphi(\varepsilon)=\varepsilon$ and $\varphi(aw)=a\varphi(w)$ if $a\in
 \Sigma\cup\Gamma$ and $\varphi((Z,\alpha)w)=Z\varphi(w)$ if $Z\in \Gamma$
 and $\alpha\in\{0,1,2\}$.
 Intuitively, $\varphi$ is a projection deleting the components in 
 $\{0,1,2\}$. 

By construction of $G_X$, if $(Z,\alpha)\to w$ is a rule of $G_X$ then
$\varphi((Z,\alpha))\to \varphi(w)$ is a rule of $G$. Therefore, if
$x_0,\ldots,x_k$ is complete derivation of $G_X$, then
$\varphi(x_0),\ldots,\varphi(x_k)$ is a complete derivation of $G$.
Moreover, the initial symbol of $G_X$ is $(S,1)$ and all rules of
$R_X$ with a left hand side in $(\Gamma\setminus\{X\})\times\{1\}$ 
have a right hand side where an element of $\Gamma\times\{1\}$ occurs.
Therefore, since $x_k\in \Sigma^*$, the only way to destroy the component $1$
is to use a rule with  the left hand side $(X,1)$. It follows that 
the derivation tree associated to $\varphi(x_0),\ldots,\varphi(x_k)$ \aloisbox{cover}s
$X$.

Consequently, $\varphi$ induces a function from $E_n(G_X)$ into
$E_{X,n}(G)$. Let $x_0,\ldots,x_k$ and $x_0^\prime,\ldots,x_k^\prime$ be
complete derivations of $G_X$, such that
$\varphi(x_0),\ldots,\varphi(x_k)=\varphi(x_0^\prime),\ldots,\varphi(x_k^\prime)$.
Assuming that  $x_0,\ldots,x_k\neq x_0^\prime,\ldots,x_k^\prime$, there
exists a minimal index $i_0$ such that $x_{i_0}\neq x_{i_0}^\prime$. 
Since $x_0=(S_0,1)=x_0^\prime$, $i_0\geq 1$. Therefore
$x_{i_0-1}=x_{i_0-1}^\prime$ exists. Set $x_{i_0-1}=v_0(Z,\alpha)v_1$, with
$Z\in \Gamma$ and $\alpha\in \{0,1,2\}$. One of the following cases arises:
\begin{itemize}
\item If $\alpha=0$, then there exist $Z\to w$ and $Z\to w^\prime$ in $R$ such that 
$x_{i_0}=v_0[w]_0v_1$ and $x_{i_0}^\prime=v_0[w^\prime]_0v_1$. Since $\varphi(x_{i_0})=\varphi(x_{i_0}^\prime)$, it follows that $\varphi([w]_0)=\varphi([w\prime]_0)$. But $\varphi([w]_0)=w$ and $\varphi([w^\prime]_0)=w^\prime$, proving that $x_{i_0}=x_{i_0}^\prime$, a contradiction.  
\item If $\alpha=2$, then the same proof holds, replacing $0$ by $2$. 
\item If $\alpha=1$ and $Z=X$, then, again, \fred{\replace{exactly }{}}the same proof \fred{\replace{as for $\alpha=0$ }{}}holds. 
\item If $\alpha=1$ and $Z\neq X$, then there exist $Z\to w$ and $Z\to
  w^\prime$ in $R$ such that $x_i=v_0w_1v_1$ and
  $x_i^\prime=v_0w_2v_1$, with $w_1\in \{w\}_{1,2}$ and $w_2\in
  \{w^\prime\}_{1,2}$.  Since
  $\varphi(x_{i_0})=\varphi(x_{i_0}^\prime)$, one has
  $w=w^\prime$. Therefore, $w_1,w_2\in \{w\}_{1,2}$.  Since $w_1\neq
  w_2$, let $j$ be the first letter of $w_1$ which is different from the
  corresponding letter in $w_2$. By construction of $\{w\}_{1,2}$,
  this letter must be in $\Gamma\times\{1,2\}$ in both $w_1$ and
  $w_2$, for instance $(T,\beta_1)$ and $(H,\beta_2)$. Now, since
  $\varphi((T,\beta_1))=\varphi((H,\beta_2))$, one has $T=H$. Therefore,
  without loss of generality we may assume that $\beta_1=1$ and
  $\beta_2=2$.  Consequently, $x_{i_0}$ has a prefix of the form
  $v_0(T,1)$: in the derivation tree corresponding to
  $x_0,\ldots,x_k$, the subtree rooted in this $(T,1)$ contains an $X$
  (by construction of $R_1$).  Conversely, $x_{i_0}^\prime$ has a
  prefix of the form $v_0(T,2)$: in the derivation tree corresponding
  to $x_0^\prime,\ldots,x_k^\prime$, the subtree rooted in this
  $(T,2)$ does not contain any $X$ (by construction of $R_2$).  It
  follows that the two corresponding derivations cannot have the same
  image by $\varphi$, a contradiction.
\end{itemize}
It follows that $\varphi$ induces an injective function from $E_n(G_X)$
into $E_{X,n}(G)$.

Now let $y_0,\ldots,y_k$ be complete derivations of $G$ whose
corresponding tree $t$ is in $E_{X,n}(G)$. We consider the tree~$t^\prime$
labelled in $\Gamma\times\{0,1,2\}\cup \Sigma$ which has
exactly the same structure (the same set of positions) than $t$ and
such that:
\begin{itemize}
\item If a node of $t$ is labelled by a letter of $\Sigma$, then the
  corresponding node in $t^\prime$ has the same label.
\item If a node $\rho$ of $t$ is labelled by a letter $T\in\Gamma$,
  then the node $\rho$ in $t^\prime$ is labelled by $(T,1)$ if there
  is no $X$ on the path from the root to $\rho$ (excluding $\rho$),
  and if the subtree rooted in $\rho$ (including $\rho$) contains one $\rho$, at
  least. It is labelled by $(T,0)$ if there is at least one
  $X$ on the path from the root to $\rho$.  Otherwise, it is labelled by
  $(T,2)$.
\end{itemize}
One can check that $t^\prime$ corresponds to a complete derivation tree
of $G_X$ whose image by $\varphi$ is exactly the complete execution
corresponding to $t$, proving that $\varphi$ is surjective, which
concludes the proof.
\end{proof}

\begin{example}[Illustration of the proof of Prop.~\ref{prop:phiX}]\label{exe:proof}
Consider the grammar $G=(\{a,b\},\{S,T,X\},S,R)$ with $R=\{S\to
SS,S\to aT,S\to Xb,T\to aa, X,\to T, X\to b\}$ of
Example~\ref{exe-GX}.  Consider the derivation tree of $E_{X,19}(G)$
depicted in Fig.~\ref{fig:exeproof}, corresponding to the complete
derivation $S, SS, SSS, aTSS, aaaSS, aaaXbS, aaabbS,
aaabbXb,\\ aaabbTXb, aaabbaaXb, aaabbaabb$. The associated derivation
in $G_X$ is $(S,1), (S,1)(S,1), (S,2)(S,1)(S,1),\\ a(T,2)(S,1)(S,1),
aaa(S,1)(S,1), aaa(X,1)b(S,1),\\ aaabb(S,1), aaabb(X,1)b,
aaabb(T,0)(X,0)b, aaabbaa(X,0)b,\\ aaabbaabb$, whose derivation tree
from $E_{19}(G_X)$ is depicted in Fig.~\ref{fig:exeproof2}.

\begin{figure}
\begin{center}
\begin{tikzpicture}
\node(0) at (0,0) {$S$};
\node(1) at (-2,-1) {$S$};
\node(2) at (2,-1) {$S$};
\node(11) at (-3,-2) {$S$};
\node(12) at (-1,-2) {$S$};
\node(111) at (-4,-3) {$a$};
\node(112) at (-3,-3) {$T$};
\node(1121) at (-3.5,-4) {$a$};
\node(1122) at (-2.5,-4) {$a$};
\node(121) at (-1,-3) {$X$};
\node(122) at (0,-3) {$b$};
\node(1211) at (-1,-4) {$b$};
\node(22) at (2,-2) {$b$};
\node(21) at (1,-2) {$X$};
\node(211) at (1,-3) {$T$};
\node(212) at (2,-3) {$X$};
\node(2111) at (0,-4) {$a$};
\node(2112) at (1,-4) {$a$};
\node(2121) at (2,-4) {$b$};

\path[] (0) edge[above] node {} (1);
\path[] (0) edge[above] node {} (2);
\path[] (1) edge[above] node {} (12);
\path[] (1) edge[above] node {} (11);
\path[] (11) edge[above] node {} (111);
\path[] (11) edge[above] node {} (112);
\path[] (112) edge[above] node {} (1121);
\path[] (112) edge[above] node {} (1122);
\path[] (12) edge[above] node {} (121);
\path[] (12) edge[above] node {} (122);
\path[] (121) edge[above] node {} (1211);
\path[] (2) edge[above] node {} (22);
\path[] (2) edge[above] node {} (21);
\path[] (21) edge[above] node {} (212);
\path[] (21) edge[above] node {} (211);
\path[] (211) edge[above] node {} (2111);
\path[] (211) edge[above] node {} (2112);
\path[] (212) edge[above] node {} (2121);
\end{tikzpicture}
\end{center}
\caption{Derivation tree of $G$ - Example~\ref{exe:proof}}\label{fig:exeproof}
\end{figure}
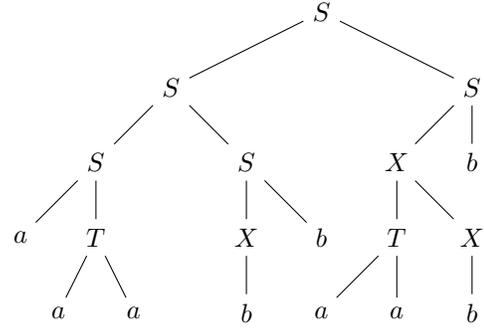

\begin{figure}
\begin{center}
\begin{tikzpicture}
\node(0) at (0,0) {$(S,1)$};
\node(1) at (-2,-1) {$(S,1)$};
\node(2) at (2,-1) {$(S,1)$};
\node(11) at (-3,-2) {$(S,2)$};
\node(12) at (-1,-2) {$(S,1)$};
\node(111) at (-4,-3) {$a$};
\node(112) at (-3,-3) {$(T,2)$};
\node(1121) at (-3.5,-4) {$a$};
\node(1122) at (-2.5,-4) {$a$};
\node(121) at (-1,-3) {$(X,1)$};
\node(122) at (0,-3) {$b$};
\node(1211) at (-1,-4) {$b$};
\node(22) at (2,-2) {$b$};
\node(21) at (1,-2) {$(X,1)$};
\node(211) at (1,-3) {$(T,0)$};
\node(212) at (2,-3) {$(X,0)$};
\node(2111) at (0,-4) {$a$};
\node(2112) at (1,-4) {$a$};
\node(2121) at (2,-4) {$b$};

\path[] (0) edge[above] node {} (1);
\path[] (0) edge[above] node {} (2);
\path[] (1) edge[above] node {} (12);
\path[] (1) edge[above] node {} (11);
\path[] (11) edge[above] node {} (111);
\path[] (11) edge[above] node {} (112);
\path[] (112) edge[above] node {} (1121);
\path[] (112) edge[above] node {} (1122);
\path[] (12) edge[above] node {} (121);
\path[] (12) edge[above] node {} (122);
\path[] (121) edge[above] node {} (1211);
\path[] (2) edge[above] node {} (22);
\path[] (2) edge[above] node {} (21);
\path[] (21) edge[above] node {} (212);
\path[] (21) edge[above] node {} (211);
\path[] (211) edge[above] node {} (2111);
\path[] (211) edge[above] node {} (2112);
\path[] (212) edge[above] node {} (2121);
\end{tikzpicture}
\end{center}
\caption{Derivation tree of $G_X$ - Example~\ref{exe:proof}}\label{fig:exeproof2}
\end{figure}
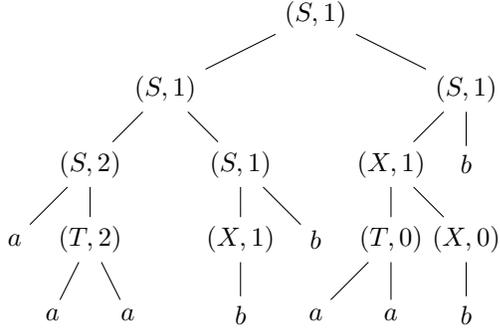
\end{example}

Using Proposition~\ref{prop:phiX} and the results described in
Section~\ref{sec-intro}, it is possible to compute $|E_{X,n}(G)|$. If we denote
by $\ell$ the maximal number of elements of $\Gamma$ (with
multiplicity) occurring in a right hand side of $G$, then $G_X$ has
$O(2^\ell |R|)$ rules, whose sizes are bounded by the maximal size of
the rules of $G$. Therefore if $\ell$ is reasonable, the computation
of $|E_{X,n}(G)|$ is tractable in practice, even for \fred{a} quite large
\fred{value of} $n$. As mentioned above, the computation of $|E_{X,n}(G)|$
immediately provides~$p_{X,n}$. It is also important to point out that
$G_X$ allows the uniform random computation of execution trees of $G$
of a given size and \aloisbox{cover}ing $X$.


Since $E_{X,X,n}(G)=E_{X,n}(G)$, computing $|E_{X,X,n}(G)|$ is a direct
application of the above techniques. Computing $|E_{X,Y,n}(G)|$, with $Y\neq
X$, can almost be done by a similar  construction: the difference is
that the construction of the rules of the grammar $G_{XY}$, from the grammar
$G_X$, must take into account that both $X$ and $Y$ have to appear in the
derivation. 
 Let $G_{XY}=(\Sigma,\Gamma\times\{0,1,2\}\times\{0,1,2\},((S_0,1),1),R_{XY})$
where $R_{XY}=R_0\cup R_1\cup R_1^\prime\cup R_2$ with:
\begin{itemize}
\item $R_0=\{((Z,i),0)\to [w]_0\mid (Z,i)\to w\in R_X\}$,
\item $R_1=\{((Z,i),1)\to w^\prime\mid Z\neq Y\text{ and }\exists (Z,i)\to w\in R_X\text{ such that }
  w^\prime\in \{w\}_{1,2}\}$,
\item $R_1^\prime=\{((Y,i),1)\to [w]_0\mid (Y,i)\to w\in R_X\}$,
\item $R_2=\{((Z,i),2)\to [w]_2\mid (Z,i)\to w\in R_X\text{ and }Z\neq Y\}$.
\end{itemize}

A proof similar to \fred{\replace{this}{the one}} of Proposition~\ref{prop:phiX} allows  showing
that there is a computable bijection between $E_{n}(G_{XY})$ and 
$E_{X,Y,n}(G)$. Note that the size of $G_{XY}$ is approximatively
$4^{\ell}$ times greater than the size of $G$.

\section{\aloisbox{Experiments}}
\label{sec-running}

The approach has been \aloisbox{evaluated} on a simplified version of the grammar
of JSON\footnote{\url{http://www.json.org/}} (for JavaScript Object Notation) -- a
 language independent  common format for declaring
objects.  Formally, let us consider the grammar  
 $G=(\Sigma,\Gamma,Object,R)$ with $\Sigma$ having the eight following elements 
$\Sigma = \{,\,, \{, \ :, \},  \ letter, \ digit, [, ]\}$.
The set $\Gamma$ of non-terminal symbols\footnote{\PCH{To provide a more readable specification, the convention
    consisting in using capital letters for non-terminal symbols is not entirely 
    respected here.}} is composed of the elements
 $''Object''$, $''Members''$,  $''Pair''$,  $''Array''$,
 $''Elements''$ and  $''Value''$. Finally, the set $R$ contains the following
rules:
\begin{itemize}
	\item $Object \to \{ \} ~|~ \{ Members \}$ 
	\item $Members \to Pair ~|~ Pair,Members$ 
	\item $Pair \to letter:Value$ 
	\item $Array \to [~ ] ~|~ [Elements]$ 
	\item $Elements \to Value ~|~ Value,Elements$ 
	\item $Value \to letter ~|~ Object ~|~ digit ~|~ Array$ 
\end{itemize}

In order to optimise the coverage criterion, we have to solve the following
system while maximising $p$ satisfying
$$
\begin{cases}
p\leq \pi_{Object} \frac{p_{Object,Object,n}}{p_{Object,n}}
+ \pi_{Members} \frac{p_{Members,Object,n}}{p_{Members,n}}
\\ \sp+ \pi_{Pair} \frac{p_{Pair,Object,n}}{p_{Pair,n}}
+ \pi_{Array} \frac{p_{Array,Object,n}}{p_{Array,n}}
\\ \sp+ \pi_{Elements} \frac{p_{Elements,Object,n}}{p_{Elements,n}}
+ \pi_{Value} \frac{p_{Value,Object,n}}{p_{Value,n}} \\

p\leq \pi_{Object} \frac{p_{Object,Members,n}}{p_{Object,n}}
+ \pi_{Members} \frac{p_{Members,Members,n}}{p_{Members,n}}
\\ \sp+ \pi_{Pair} \frac{p_{Pair,Members,n}}{p_{Pair,n}}
+ \pi_{Array} \frac{p_{Array,Members,n}}{p_{Array,n}}
\\ \sp+ \pi_{Elements} \frac{p_{Elements,Members,n}}{p_{Elements,n}}
+ \pi_{Value} \frac{p_{Value,Members,n}}{p_{Value,n}} \\

p\leq \pi_{Object} \frac{p_{Object,Pair,n}}{p_{Object,n}}
+ \pi_{Members} \frac{p_{Members,Pair,n}}{p_{Members,n}}
\\ \sp+ \pi_{Pair} \frac{p_{Pair,Pair,n}}{p_{Pair,n}}
+ \pi_{Array} \frac{p_{Array,Pair,n}}{p_{Array,n}}
\\ \sp+ \pi_{Elements} \frac{p_{Elements,Pair,n}}{p_{Elements,n}}
+ \pi_{Value} \frac{p_{Value,Pair,n}}{p_{Value,n}} \\

p\leq \pi_{Object} \frac{p_{Object,Array,n}}{p_{Object,n}}
+ \pi_{Members} \frac{p_{Members,Array,n}}{p_{Members,n}}
\\ \sp+ \pi_{Pair} \frac{p_{Pair,Array,n}}{p_{Pair,n}}
+ \pi_{Array} \frac{p_{Array,Array,n}}{p_{Array,n}}
\\ \sp+ \pi_{Elements} \frac{p_{Elements,Array,n}}{p_{Elements,n}}
+ \pi_{Value} \frac{p_{Value,Array,n}}{p_{Value,n}} \\

p\leq \pi_{Object} \frac{p_{Object,Elements,n}}{p_{Object,n}}
+ \pi_{Members} \frac{p_{Members,Elements,n}}{p_{Members,n}}
\\ \sp+ \pi_{Pair} \frac{p_{Pair,Elements,n}}{p_{Pair,n}}
+ \pi_{Array} \frac{p_{Array,Elements,n}}{p_{Array,n}}
\\ \sp+ \pi_{Elements} \frac{p_{Elements,Elements,n}}{p_{Elements,n}}
+ \pi_{Value} \frac{p_{Value,Elements,n}}{p_{Value,n}} \\

p\leq \pi_{Object} \frac{p_{Object,Value,n}}{p_{Object,n}}
+ \pi_{Members} \frac{p_{Members,Value,n}}{p_{Members,n}}
\\ \sp+ \pi_{Pair} \frac{p_{Pair,Value,n}}{p_{Pair,n}}
+ \pi_{Array} \frac{p_{Array,Value,n}}{p_{Array,n}}
\\ \sp+ \pi_{Elements} \frac{p_{Elements,Value,n}}{p_{Elements,n}}
+ \pi_{Value} \frac{p_{Value,Value,n}}{p_{Value,n}} \\

\pi_{Object} + \pi_{Members} + \pi_{Pair} + \pi_{Array} + \pi_{Elements} + \pi_{Value} =1
\end{cases}
$$
 
Using a \fred{\replace{lightly}{slightly}} modified version of the Hoa tool (\cite{enderlin_hoa}), the
computation of  the probabilities $p_{X,n}$ and $p_{X,Y,n}$ for all $X, Y \in
\Gamma$ and $n=20$ has been performed efficiently \aloisbox{(a few seconds)}. 
The system becomes as below, and we have then  
to solve it while maximising $p$ satisfying
$$
\begin{cases}
p\leq \pi_{Object} \frac{12}{12}
+ \pi_{Members} \frac{12}{12}
+ \pi_{Pair} \frac{12}{12}
+ \pi_{Array} \frac{11}{11}
\\ \sp+ \pi_{Elements} \frac{8}{8}
+ \pi_{Value} \frac{12}{12} \\

p\leq \pi_{Object} \frac{12}{12}
+ \pi_{Members} \frac{12}{12}
+ \pi_{Pair} \frac{12}{12}
+ \pi_{Array} \frac{11}{11}
\\ \sp+ \pi_{Elements} \frac{8}{8}
+ \pi_{Value} \frac{12}{12} \\

p\leq \pi_{Object} \frac{12}{12}
+ \pi_{Members} \frac{12}{12}
+ \pi_{Pair} \frac{12}{12}
+ \pi_{Array} \frac{11}{11}
\\ \sp+ \pi_{Elements} \frac{8}{8}
+ \pi_{Value} \frac{12}{12} \\

p\leq \pi_{Object} \frac{11}{12}
+ \pi_{Members} \frac{11}{12}
+ \pi_{Pair} \frac{11}{12}
+ \pi_{Array} \frac{11}{11}
\\ \sp+ \pi_{Elements} \frac{8}{8}
+ \pi_{Value} \frac{11}{12} \\

p\leq \pi_{Object} \frac{8}{12}
+ \pi_{Members} \frac{8}{12}
+ \pi_{Pair} \frac{8}{12}
+ \pi_{Array} \frac{8}{11}
\\ \sp+ \pi_{Elements} \frac{8}{8}
+ \pi_{Value} \frac{8}{12} \\

p\leq \pi_{Object} \frac{12}{12}
+ \pi_{Members} \frac{12}{12}
+ \pi_{Pair} \frac{12}{12}
+ \pi_{Array} \frac{11}{11}
\\ \sp+ \pi_{Elements} \frac{8}{8}
+ \pi_{Value} \frac{12}{12} \\

\pi_{Object}
 + \pi_{Members}
 + \pi_{Pair}
 + \pi_{Array}
\\ \sp + \pi_{Elements}
 + \pi_{Value} =1
\end{cases}
$$

This linear programming problem can be solved in an efficient way, using
simplex-like approaches. We have used the tool
lp\_solve\footnote{\url{http://lpsolve.sourceforge.net/}} to solve it, and
the result is that $p=1$ if $\pi_{Object}=0$, $\pi_{Members}=0$,
$\pi_{Pair}=0$, $\pi_{Array}=0$, $\pi_{Elements}=1$, and $\pi_{Value}=0$. It
means that, for this simple example, the optimised approach \aloisbox{to cover all the non-terminals symbols,} consists in
generating derivation tree\aloisbox{s} \PCH{covering} $Elements$.
\OK{Indeed, }\aloisbox{in this grammar, the generation of \PCH{a derivation
    tree covering} the \OK{non-terminal} symbol $Elements$ 
\PCH{provides a tree covering all the other non-terminal symbols}}.
\section{Conclusion}\label{sec-conclusion}

In this paper\fred{,} we have presented a method for exploiting a coverage
criterion together with random testing in the
\fred{\replace{grammar-based testing context}{context of grammar-based testing}}.
This automatic method lies in building a grammar and then in
resolving a linear constraint system, which can be done by adapted tools,
even for large values. In the future\fred{,} we plan to extend the
approach to other coverage criteria \fred{\replace{like covering rules}{such as rules coverage}},  and
\OK{also to handle attribute grammars with constraints
formalising the semantics of context-free languages}.

\bibliographystyle{IEEEtrans}
\bibliography{mbtalea}


\end{document}